\documentclass[twocolumn,aps,showpacs,prb,tightenlines,amsmath,amssymb,superscriptaddress]{revtex4}
\usepackage{graphicx}
\usepackage{amssymb}
\usepackage{amsmath}
\usepackage{bm}

\usepackage{colordvi} 
\newcommand{\bgreek}[1]{\mbox{\boldmath$#1$\unboldmath}}
\begin{document} 

\title{Hole spin relaxation in intrinsic and $p$-type bulk GaAs}

\author{K.\ Shen}
\affiliation{Hefei National Laboratory for Physical Sciences at
  Microscale, 
University of Science and Technology of China, Hefei,
  Anhui, 230026, China}
\author{M.\ W.\ Wu}
\thanks{Author to whom correspondence should be addressed}
\email{mwwu@ustc.edu.cn.}
\affiliation{Hefei National Laboratory for Physical Sciences at
Microscale,
University of Science and Technology of China, Hefei,
Anhui, 230026, China}
\affiliation{Department of Physics, 
University of Science and Technology of China, Hefei,
Anhui, 230026, China}

\date{\today}

\begin{abstract}
We investigate hole spin relaxation in
intrinsic and $p$-type bulk GaAs from a
fully microscopic kinetic spin Bloch equation approach. In contrast to the
previous study on hole spin dynamics, we explicitly
include the intraband coherence and the nonpolar
hole-optical-phonon interaction, both of which are demonstrated to be
of great importance to the hole spin relaxation. The relative
contributions of the D'yakonov-Perel' and 
Elliott-Yafet mechanisms on hole spin relaxation are also analyzed. In
our calculation, the screening constant, playing an important role 
in the hole spin relaxation, is treated with the random
phase approximation. In intrinsic GaAs, our result
shows good agreement with the experiment data at room
temperature, where the hole spin relaxation is demonstrated to be
dominated by the Elliott-Yafet mechanism.
We also find that the hole spin relaxation strongly depends on the
temperature and predict a valley in the density dependence of the hole
spin relaxation time at low temperature due to the hole-electron scattering. In
$p$-type GaAs, we predict a peak in the spin relaxation time against
the hole density at low temperature,
which originates from the distinct behaviors of the screening
in the degenerate and nondegenerate regimes. The competition between
the screening and
the momentum exchange during scattering events can also lead to
a valley in the density dependence of the hole spin relaxation time in
the low density regime. At high temperature,
the effect of the screening is suppressed due to the
small screening constant.
Moreover, we predict a nonmonotonic dependence
of the hole spin relaxation time on temperature associated with the
screening together with the hole-phonon scattering. Finally, we find that the
D'yakonov-Perel' mechanism can markedly contribute to the
hole spin relaxation in the low density case at moderate temperature
and in the high density case at low temperature, where the Elliott-Yafet
mechanism is suppressed due to the relatively weak scattering.
\end{abstract}

\maketitle

\section{Introduction}
\label{introduction}

In the past decade, semiconductor spintronics,
with the aim of realizing favourable devices for future application
based on the spin degree of 
freedom, has attracted much attention.\cite{meier,wolf,spintronics,korn,wu}
One of the most critical challenges for such devices lies in the
control of the spin lifetime, which is limited by the unavoidable spin
relaxation and/or dephasing process in semiconductors. In this sense,
understanding of the carrier spin relaxations and/or dephasings is a
critical issue.\cite{spintronics,korn,wu,chen,sherman} In bulk materials, the
spin relaxation and/or dephasing properties of the electrons 
and the underlying physics have been well understood after a long-time
research.\cite{seymour,zerro,kikkawa,song,teng,oertel,wu6,romer,krauss2,shen}
However, the study on hole spin
dynamics, which occurs on the time scale of the momentum scattering time
(usually $<1$~ps), is fairly rare, partly because of the limited resolution
on the detection of such an ultrafast process. To our
best knowledge, the hole spin lifetime in bulk semiconductors was only 
measured in intrinsic GaAs at room temperature, which evaluated the
spin relaxation time
of the heavy hole ($\sim 110$~fs).\cite{hilton}

One picture to explain the picosecond hole spin relaxation (HSR) time is
the D'yakonov-Perel'-like description associated with the momentum scattering
and the spin precession between the light-hole (LH) and heavy-hole
(HH).\cite{culcer} However, this picture works only if 
the broadening of the energy spectrum due to scattering
is larger than the {\it interband} 
splitting, corresponding to the vicinity of the zone center. Otherwise, holes 
are driven to large momentum states where the degeneracy between
the LH and HH bands
is significantly lifted, hence it is more appropriate to treat the LH
and HH bands separately. The HSR of the individual LH
and HH bands was
studied by Yu {\em et al.},\cite{yu} who obtained the
accurate band structure from the tight-binding Hamiltonian with
the spin-orbit coupling (SOC) and calculated the spin relaxation time
of the LH (HH) as the inversion of
the decay rate of the quasi-spin polarization, that is,
  the population difference of the LH (HH) between the two quasi-spin bands.
The Elliott-Yafet (EY)
mechanism,\cite{EY} there described as the direct quasi-spin-flip 
scattering, was claimed to be the solo mechanism for the HH and the
dominant one for the LH, while the D'yakonov-Perel' (DP)
mechanism\cite{DP} treated as the intraband precession together with the quasi-spin-conserving
scattering from the motional narrowing relation was found to be
unimportant. Since the calculation was based on the single particle
theory, the Coulomb interaction which has been demonstrated to result in
intriguing many-body effects during electron spin dynamics\cite{wu,wu1,wu2,glazov} was
missed in that work. Moreover, the HSR time was extracted
from the quasi-spin polarization instead of the exact spin signal.
The feasibility of this approach needs to be verified. 
In other words, the results obtained there is not exactly the HSR time,
  but the quasi-spin relaxation time.
Recently, the hole spin
dynamics in intrinsic bulk GaAs with the Coulomb interaction was
microscopically investigated from an eight-band Kane
  Hamiltonian by Krau{\ss} {\em et 
  al.},\cite{krauss} where the HSR
  time, directly from the decay of the hole spin
  expectation, was shown to be quite close to the 
  experimental result.
It was also shown that the HSR time can
be slightly different from the quasi-spin relaxation time. Thus, it seems
that ultrafast HSR in bulk zinc-blende
semiconductors has been successfully interpreted 
theoretically. However, one may notice that the {\it intraband} coherence, i.e., the
non-diagonal components of the density matrices was missing
in that work, which actually will lead to two consequences:
the exaggeration of the EY mechanism and the
exclusion of the DP mechanism. Moreover, the nonpolar interaction of holes
with transverse- and longitudinal-optical-phonons due to the deformation potential 
coupling, which was shown to produce a significant
contribution on the charge dynamics of holes,\cite{langot,scholz,brudevoll} was not
included. In this sense, the results in Ref.\,\onlinecite{krauss}
should be reexamined.

In the present work, we employ the fully
microscopic kinetic spin Bloch equations (KSBEs)\cite{wu,wu1,wu6,wu9,wu10} to
investigate the HSR due
to the EY and DP mechanisms in intrinsic and $p$-type bulk GaAs. 
The EY mechanism here corresponds to the decay of the spin polarization
solely due to the scattering associated with the interband mixing,
between not only the HH and LH bands but also the conduction and
valence bands; whereas the DP 
mechanism stands for the additional
contribution by including the intraband spin precessions.
We first analytically derive the valence band structure 
from the four-band Luttinger Hamiltonian\cite{luttinger} together with
the Dresselhaus SOC (Ref.\,\onlinecite{dressel}) due to the bulk
inversion asymmetry in zinc-blende semiconductors. We demonstrate that the
Dresselhaus SOC induces an
intraband splitting between the two HH bands as well as the two LH bands
with cubic wave-vector dependence,
which supplies an additional HSR channel due to the DP
mechanism. We find that the 
intraband splitting of the HH bands vanishes under the spherical
approximation of the Luttinger Hamiltonian, and
then the DP mechanism becomes irrelevant to the HH spin
relaxation.\cite{yu} This implies the limitation of such
an approximation in calculating the HSR time. Therefore, we later obtain the intraband
splitting and wave functions by diagonalizing the full
eight-band ${\bf k}\cdot {\bf p}$ Hamiltonian {\it beyond} the spherical
approximation. Then we calculate the 
HSR time from the KSBEs with the {\it intraband} coherence
and all the relevant scatterings, such as the hole-impurity,
hole-hole, hole-electron, hole-acoustic-phonon, polar and
nonpolar hole-optical-phonon scatterings, explicitly 
included. In the intrinsic case at room temperature, the HSR is
dominated by the EY mechanism and the HSR time shows
good agreement with the experiment. We find that the 
HSR time can be significantly manipulated by changing the
temperature and excitation density.
In the $p$-type materials, we predict intriguing nonmonotonic
behaviors of the HSR time in both the density and
temperature dependences. We show that the nonmonotonic features
reflect the role of the screening. Moreover, we find that the EY
mechanism is usually the major mechanism of the HSR,
but the DP mechanism can still be comparable with the EY mechanism in
some special cases, such as the high density regime at low temperature
and the low density regime at moderate temperature.

This paper is organized as follows. In Sec.~II, we set up our model and
derive the effective Dresselhaus field from the Luttinger
Hamiltonian. The KSBEs are also constructed in this section. In Sec.~III,
we investigate the HSR in both intrinsic and $p$-type
GaAs. The comparison of the calculations with and
without the intraband coherence is
given in this section to illustrate the role of 
the intraband coherence. The relative contributions of the DP and EY
mechanisms are also discussed. Finally, we summarize in Sec.~IV.

\section{Model and KSBE}
\label{model}

To qualitatively analyze the band structure and the intraband splitting of the HH
and LH bands, we start from the perturbation method 
with the $4\times 4$ Hamiltonian (in the basis of the eigenstates
of $J_z$ with eigenvalues $\tfrac{3}{2}$, $\tfrac{1}{2}$,
$-\tfrac{1}{2}$, and $-\tfrac{3}{2}$, in sequence) near the center of the Brillouin-zone
\begin{eqnarray}
{\mathcal H}_{8v8v}=\left(
\begin{array}{cccc}
F& H&I&0\\
H^\ast&G&0&I\\
I^\ast&0&G&-H\\
0&I^\ast&-H^\ast &F
\end{array}
\right)+H_{8v8v}^b.
\label{eq1}
\end{eqnarray}
The first term on the right-hand side of the equation is the Luttinger
Hamiltonian,\cite{luttinger}
where $F=-\frac{\hbar^2}{2m_0}[(\gamma_1+\gamma_2)(k_x^2+k_y^2)+(\gamma_1-2\gamma_2)k_z^2]$,
$G=-\frac{\hbar^2}{2m_0}[(\gamma_1-\gamma_2)(k_x^2+k_y^2)+(\gamma_1+2\gamma_2)k_z^2]$,
$H=2\sqrt{3}\frac{\hbar^2}{2m_0}\gamma_3(k_x-ik_y)k_z$, and
$I=\frac{\hbar^2}{2m_0}[\sqrt{3}\gamma_2(k_x^2-k_y^2)-i2\sqrt{3}\gamma_3k_xk_y]$. $\gamma_i$
are Kohn-Luttinger parameters. We denote this term as $H_0$ in the
following. The second term is the Dresselhaus SOC of the valence band, $H_{8v8v}^b={\bf
  h}_{\bf k}\cdot {\bf J}$, with the effective magnetic field\cite{dressel,winkler}
\begin{equation}
{\bf h}_{\bf k}=b_{41}^{8v8v}[k_x(k_y^2-k_z^2), k_y(k_z^2-k_x^2),
k_z(k_x^2-k_y^2)].
\label{eq2}
\end{equation}
$J_i$ represent the spin-3/2 angular-momentum matrices. It is
obvious that the band structure is mainly determined by $H_0$ in the
vicinity of the zone-center. The energy spectrum from
$H_0$ reads\cite{dressel2,chao}
\begin{eqnarray}
  \nonumber
  E_{h/l,\bf k}&=&-\tfrac{\hbar^2}{2m_0}\big\{\gamma_1 k^2\mp[4\gamma^2_2k^4
  +12(\gamma_3^2-\gamma_2^2)\\
  &&\mbox{}\times
  (k_x^2k_y^2+k_y^2k_z^2+k_z^2k_x^2)]^{\tfrac{1}{2}}\big\},
  \label{eq3}
\end{eqnarray}
and the wave functions can be expressed by
\begin{eqnarray}
\label{eq4}
  \psi^{h/l}_1=(a_1^{h/l},b_1^{h/l},c_1^{h/l},0)^T,\\
  \psi^{h/l}_2=(0,c_2^{h/l},b_2^{h/l},a_2^{h/l})^T.
  \label{eq4-5}
\end{eqnarray}

By identifying $\psi_1^{h/l}$ and $\psi_2^{h/l}$
  as pseudo-spin-up and -down states
  separately, one obtains the effective 
magnetic field of the Dresselhaus SOC (${\bgreek \Omega}^e_{h/l}$) between them.
Specifically, one rewrites $H_{8v8v}$ in the representation of
$\{\psi_i^{h/l}\}$, and derives individual HH and LH blocks from the
L\"owdin partitioning method\cite{lowdin} upto 
 the cubic power of the wave vector,
\begin{equation}
  H_{h/l}({\bf k})=E_{h/l,{\bf k}}I_{2\times 2}+{\bgreek
    \Omega}^e_{h/l}({\bf k})\cdot{\bgreek \sigma},
\label{eq6}
\end{equation}
where $\bgreek \sigma$ are the Pauli matrices. The expressions of the effective
magnetic field together with the coefficients
$a_i^{h/l}$, $b_i^{h/l}$ and $c_i^{h/l}$ are given in
Appendix\,\ref{appen}. The presence of the intraband splitting $\Delta
E_{h/l,\bf k}=E^2_{h/l,\bf 
    k}-E^1_{h/l,\bf k}=2|{\bgreek
    \Omega}^e_{h/l}({\bf k})|$ obviously
  indicates that the DP mechanism can contribute to the HSR in the
  presence of the scatterings.\cite{wu,DP} Interestingly, one finds ${\bgreek
  \Omega}^e_{h}$ vanishes once the spherical approximation
(corresponding to $\gamma_2=\gamma_3$)\cite{balde,schliemann} is applied,
which means that the contribution
of the DP mechanism to the HH spin relaxation is lost within
the spherical approximation scheme. In other words, the difference between $\gamma_2$ and
$\gamma_3$, arising from the remote bands,\cite{winkler} is important
in counting the HH spin relaxation.

We should point out that the above perturbation approach
gives the precise value of the intraband splitting only in the vicinity 
of the zone-center. For the regime far away from the center,
the modification from the split-off and conduction bands should be
considered. Therefore, the intraband splittings $\Delta E_{h/l,\bf k}$
are obtained from the diagonalization
of the $8\times 8$ Kane Hamiltonian $H_K$ (Ref.\,\onlinecite{kane}) in our numerical
calculation, that is, by solving the Schr\"odinger equation
\begin{equation}
H_K({\bf k})|\xi_{h/l,\bf k}^i\rangle=E^i_{h/l,{\bf k}}|\xi_{h/l,\bf k}^i\rangle,
\label{eq7}
\end{equation}
with $i=1,2$. We define the single particle density matrices $\rho_{\bf k}^h$
as $4\times 4$ matrices in the helix
representation,\cite{wu4} i.e., under the basis of the
  eigenstates
$\{|\xi_{h,\bf k}^1\rangle,|\xi_{h,\bf k}^2\rangle,|\xi_{l,\bf
  k}^1\rangle,|\xi_{l,\bf k}^2\rangle\}$. The unitary transformation
from the helix representation to the
  collinear representation, the basis of which is
    defined as the eigenstates of the angular momentum operator $J_z$, is given by
\begin{equation}\rho_{\bf k}^c=U_{\bf k}\rho_{\bf k}^hU_{\bf
  k}^\dag,
\label{eq8}
\end{equation}
with $U_{\bf k}=(\xi_{h,\bf k}^1,\xi_{h,\bf k}^2,\xi_{l,\bf
    k}^1,\xi_{l,\bf k}^2)$.
The KSBEs in the helix representation from the nonequilibrium Green-function
method reads\cite{wu,wu1}
\begin{equation}
\partial_t \rho^h_{\bf k}=\partial_t \rho^h_{\bf
      k}\big|_{\rm coh}+\partial_t\rho^h_{\bf
      k}\big|_{\rm scat}.
\label{eq9}
\end{equation}
The coherent term can be written as
\begin{eqnarray}
  \nonumber
  \partial_t \rho^h_{{\bf k}}\big|_{\rm
    coh}&=&-i\big[\sum_{{\bf q}}V_{\bf q}S_{\bf k,k-q} 
  \rho^h_{\bf k-q}S_{\bf k-q,k},\rho^h_{\bf k}\big]\\
  &&\mbox{}-i\big[H^h({\bf k}), \rho^h_{\bf k}\big],
\label{eq10}
\end{eqnarray}
with $[,]$ representing the commutator. $S_{\bf
  k^\prime,k}=U_{\bf k^\prime}^\dag U_{\bf k}$.
The first term on the right-hand side of the equation comes from the Coulomb
Hartree-Fock contribution, which can be
neglected for small spin polarization.\cite{wu} In the helix frame,
$H^h({\bf k})={\rm diag}(E_{h,\bf k}^1,E_{h,\bf k}^2,E_{l,\bf
  k}^1,E_{l,\bf k}^2)$, and Eq.\,(\ref{eq10})
then can be rewritten as $\partial_t\rho_{\bf k}^h(m,n)\big|_{\rm
  coh}=-i\rho_{\bf k}^h(m,n)(E_{m,{\bf k}}-E_{n,{\bf
    k}})$. The scattering term is given by\cite{wu4}
\begin{gather}
\nonumber\hspace{-1cm}
\shoveleft\partial_t\rho_{\bf k}^h\big|_{\rm scat}=-\pi n_i\sum_{{\bf
      q}\eta\eta^\prime}|U^i_{\bf q}|^2\delta(E_{\eta^\prime{\bf
      k-q}}-E_{\eta{\bf k}})S_{\bf k,k-q}\\
\nonumber
   \mbox{} \times\big[\rho_{\bf k-q}^{h,>}T_{\bf k-q}^{\eta^\prime}S_{\bf
    k-q,k}T_{\bf k}^{\eta}\rho_{\bf
    k}^{h,<}-\rho_{\bf k-q}^{h,<}
  T_{\bf k-q}^{\eta^\prime}S_{\bf
    k-q,k}T_{\bf k}^{\eta}\rho_{\bf k}^{h,>}\big]\\
  \nonumber
  \mbox{}-\pi \sum_{{\bf
      q}\eta\eta^\prime\lambda}|M_{{\bf q},\lambda}|^2\delta(E_{\eta^\prime{\bf
      k-q}}-E_{\eta{\bf k}}\pm\omega_{{\bf q}\lambda})S_{\bf
    k,k-q}\\
\nonumber
   \mbox{}\times \big[N_{{\bf q},\lambda}^\pm\rho_{\bf k-q}^{h,>}T_{\bf k-q}^{\eta^\prime}S_{\bf
    k-q,k}T_{\bf k}^{\eta}\rho_{\eta\bf k}^{h,<}\\
  \nonumber
  \mbox{}-N_{{\bf q},\lambda}^\mp\rho_{\bf k-q}^{h,<}T_{\bf k-q}^{\eta^\prime}S_{\bf
    k-q,k}T_{\bf k}^{\eta}\rho_{\bf k}^{h,>}\big]-\pi \sum_{{\bf q}\eta^\prime}V^2_{\bf q}\\
  \nonumber
\mbox{}\times\big[Q({\bf
  q},E_{\eta{\bf k}}-E_{\eta^\prime{\bf k-q}})S_{\bf
  k,k-q}\rho^{h,>}_{\bf k-q}T_{\bf k-q}^{\eta^\prime}S_{\bf
    k-q,k}T_{\bf k}^{\eta}\rho^{h,<}_{\bf k} \\
\nonumber
 \mbox{}-Q({\bf -q},E_{\eta^\prime{\bf
      k-q}}-E_{\eta\bf k})
S_{\bf
    k,k-q}\rho^{h,<}_{\bf k-q}T_{\bf k-q}^{\eta^\prime}S_{\bf
    k-q,k}T_{\bf k}^{\eta}\rho^{h,>}_{\bf k}\big]\\
  \mbox{}+{\rm H.c.},
  \label{eq11}
\end{gather}
where, $\rho^{h,<}_{\bf k}=\rho^{h}_{\bf k}$ and
$\rho^{h,>}_{\bf k}=1-\rho^{h}_{\bf k}$. $T_{\bf k}^{\eta}(m,n)=\delta_{m,\eta}\delta_{n,\eta}$
The hole-impurity scattering matrix element $|U_{\bf
  q}^i|^2=Z^2V_{\bf q}^2$ with $Z$ taken to be 1 in the
calculation. $V_{\bf q}=e^2/[\kappa_0(q^2+\kappa^2)]$.
$\kappa_0$ denotes the 
static dielectric constant. Here, the screening constant $\kappa$
is calculated from the random phase approximation
(RPA).\cite{schliemann} 
The detail of the polar
carrier-longitudinal-optical(LO)-phonon and carrier-acoustic(AC)-phonon scattering elements
$|M_{{\bf q},\lambda}|^2$ can be 
found in Refs.\,\onlinecite{wu2} and \onlinecite{wu3}. Besides,
the longitudinal and transverse optical modes can also contribute to the nonpolar
hole-optical-phonon scattering. The matrix elements of these scatterings are
given\cite{scholz} by $M_{{\bf 
    q},\lambda}^{\rm nonp}=(\tfrac{\hbar}{2D\omega_{{\bf
      q},\lambda}})^{1/2}{\bar D}_{{\bf q},\lambda}$, where $D$ in the
  square root represents the crystal density. The potential
matrix ${\bar D}_{{\bf q},\lambda}$ is given in
Appendix\,\ref{nonpolar}. $N_{{\bf q},\lambda}^\pm =[{\rm 
  exp}(\omega_{{\bf
    q},\lambda}/k_BT)-1]^{-1}+\tfrac{1}{2}\pm\tfrac{1}{2}$. The
function $Q$ in the Coulomb scattering term reads
\begin{eqnarray}
\nonumber
&&\hspace{-1cm}Q({\bf q},w)=\sum_{\eta\eta^\prime \bf k^{\prime\prime}}\delta(E_{\eta\bf
    k^{\prime\prime}}-E_{\eta^\prime\bf k^{\prime\prime}- q}-w) \\
  \nonumber
&&\mbox{}\times  {\rm Tr}[S_{\bf
    k^{\prime\prime}-q,k^{\prime\prime}}\rho^{h,>}_{\bf
    k^{\prime\prime}}T_{\bf k^{\prime\prime}}^{\eta}S_{\bf
    k^{\prime\prime},k^{\prime\prime}-q}T_{\bf k^{\prime\prime}-q}^{\eta^\prime}\rho^{h,<}_{\bf
    k^{\prime\prime}- q}]\\
&&  \mbox{}+\sum_{\bf k^{\prime\prime}}\delta(E_{e\bf
    k^{\prime\prime}}-E_{e\bf k^{\prime\prime}- q}-w){\rm Tr}[\rho^{>}_{e,\bf
    k^{\prime\prime}}\rho^{<}_{e,\bf
    k^{\prime\prime}- q}].
  \label{eq12}
\end{eqnarray}
The second term on the right-hand side of the above equation describes the
  contribution of the hole-electron scattering with $\rho_{e,\bf
    k}$ representing the electron density matrices.

\section{Numerical Results}\label{results}

\begin{figure}[bth]
\includegraphics[width=6cm]{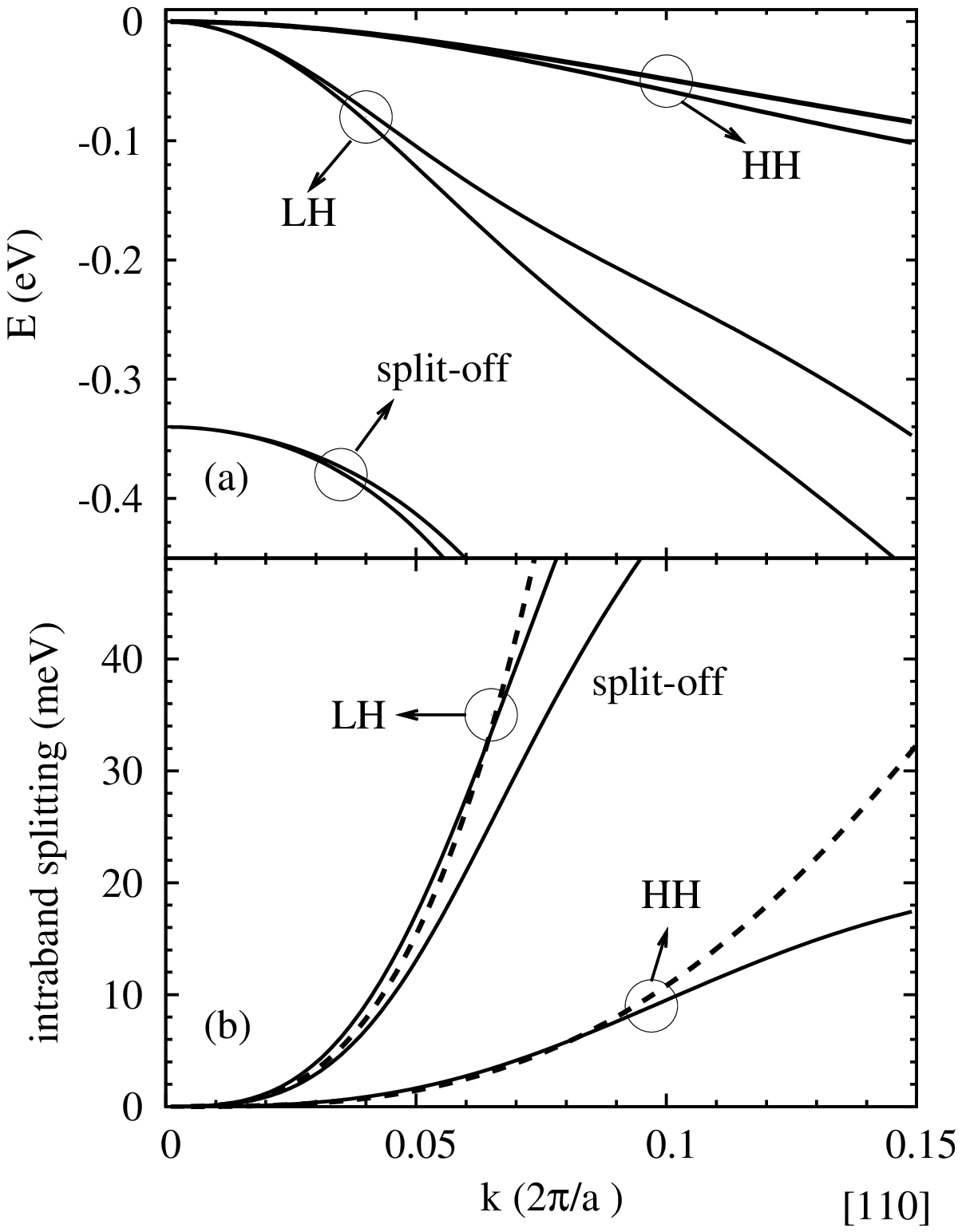}
\includegraphics[width=8.cm]{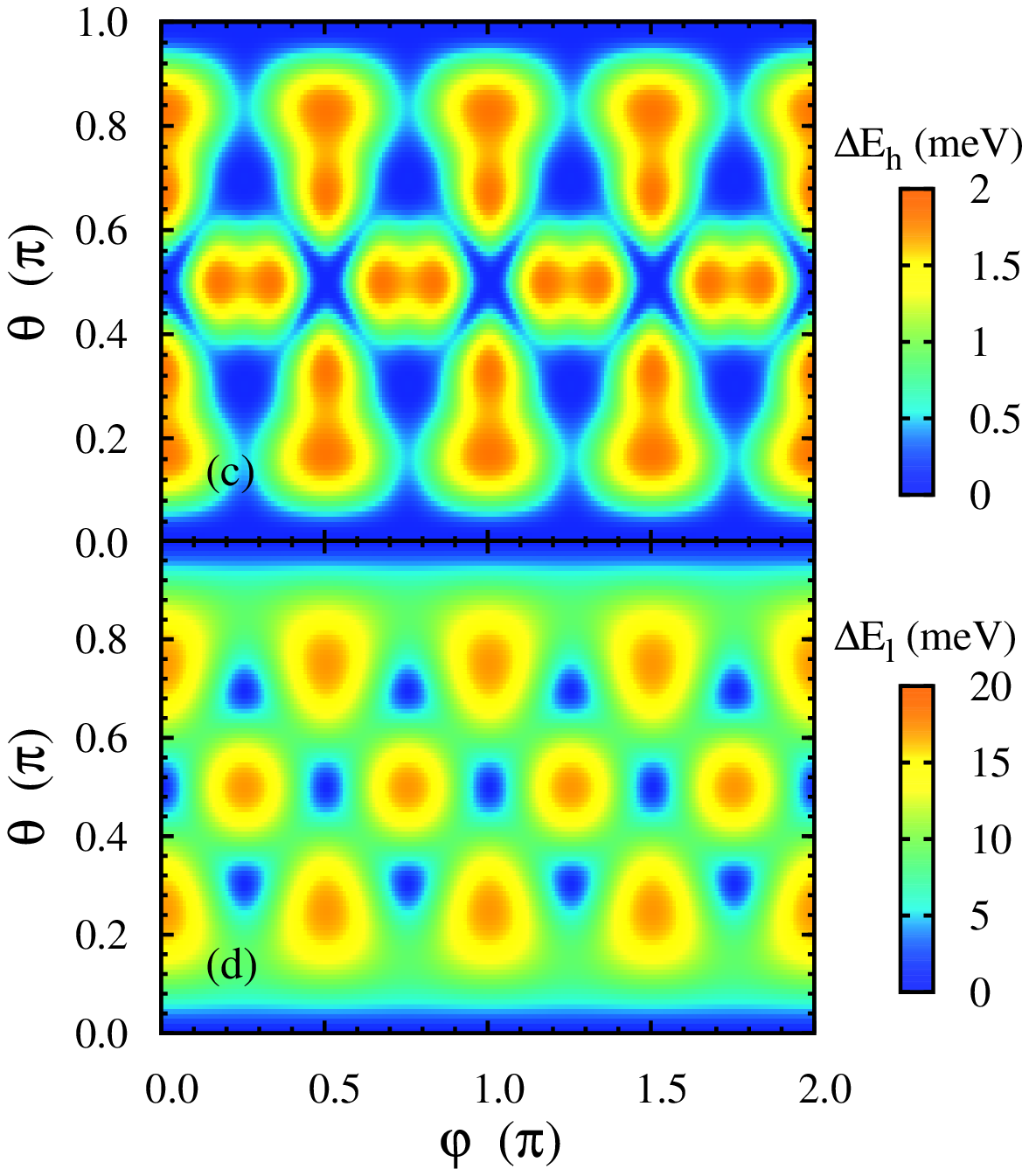}
\caption{(Color online) (a) Energy spectra and (b) intraband
  splittings of the valence bands (solid curves) along the
  $[110]$-direction from the diagonalization of the
  $8\times 8$ Kane
  Hamiltonian. The dashed curves in (b)
  show the results from the $4\times 4$ Luttinger Hamiltonian. $a$ is the 
  lattice constant. (c) and (d) show the intraband splittings of the HH and LH bands at
  $|{\bf k}|=0.1\pi/a$ as
  function of the wave direction $(\theta,\psi)$.} 
\label{fig1}
\end{figure}
In this section, we present our results in bulk GaAs with the measured
value of the optical deformation potential
$d_0=48$~eV.\cite{grim,langot} The other parameters\cite{wu6} in our
computation are all taken from
Ref.\,\onlinecite{madelung}. By numerically diagonalizing
the $8\times 8$ Kane Hamiltonian, one obtains the energy spectra of the
valence bands. The results along the $[110]$-direction are 
plotted in Fig.\,\ref{fig1}(a), where the HH, LH and split-off bands
all present intraband splittings. One notices that both the HH and
LH bands are approximately parabolic within $0.15$~eV from the valence-band top, which indicates
the feasibility of the effective-mass approximation. Moreover, the
splitting between the HH and LH bands are tens of meV even 
for the Fermi energy smaller than $10$~meV, corresponding to
$n_h\approx 3\times 10^{16}$~cm$^{-3}$ at 0~K. Therefore, we
neglect the interband coherence between the HH and
LH bands and reduce the $4\times 4$ hole density matrices into the LH and
HH blocks, both are $2\times
2$ matrices, in our computation.\cite{com_spliting} In 
Fig.\,\ref{fig1}(b), we illustrate the intraband splittings of the three valence bands, i.e., $\Delta
E_{\eta,\bf k}=E_{\eta,\bf k}^2-E_{\eta,\bf k}^1$, from the $8\times
8$ Kane Hamiltonian along the $[110]$-direction as solid
curves. Here, $E_{\eta}^2$ ($E_{\eta}^1$) is the larger (smaller)
eigenvalue of the $\eta$-band with $\eta$ corresponding to the HH, LH
and split-off bands. One can see that the splitting of the HH band is much smaller
than that of the LH band. This can be understood as
follows: Within the spherical approximation, the LH band itself presents an
intraband splitting due to the Dresselhaus 
SOC, while the HH band is
still doubly degenerate as mentioned in Sec.~II. However, the
anisotropy property of the valence band makes the
HH states contain some LH components so as to lift the degeneracy of
the HH band. In other words, the SOC induces the 
splitting of the HH band indirectly, hence the magnitude is smaller than the direct splitting
of the LH band. For comparison, we also plot the intraband
splittings from the
perturbation approach up to the order of $k^3$ based on the $4\times 4$
Luttinger Hamiltonian in Fig.\,\ref{fig1}(b) as dashed curves. One
can see that the perturbation approach
only performs well for $k<0.16\pi/a$. For larger wave vectors, the
cubic wave-vector dependence of the intraband splitting
is violated due to the interband coupling. The intraband splittings of
the HH and LH bands are plotted as function of the direction of the
wave vector in Fig.\,\ref{fig1}(c) and (d), respectively, from which
the anisotropy properties can be clearly seen.

Since the intraband splitting is much smaller in the density and temperature
regimes studied in the present work compared to the Fermi energy,
one can neglect the intraband splitting in the $\delta$-function in
 Eqs.\,(\ref{eq11}) and (\ref{eq12}).\cite{wu4} For the
 numerical treatment of the scattering term, we take the isotropic
 energy spectrum from the effective-mass approximation, $E^1_{h/l,{\bf
      k}}\approx E^2_{h/l,{\bf
      k}}\approx E_{h/l,{\bf k}}=
  -\tfrac{\hbar^2}{2m_0}(\gamma_1\mp\bar{\gamma})k^2$ with
  $\bar\gamma=(\gamma_2+\gamma_3)/2$.\cite{schliemann,wu6} 
The feasibility of this widely adopted approximation has been shown in the
  literature.\cite{sliwa,sham,csontos} Moreover, we find that the density
  of states from this approximation is almost the same as that
  obtained from the anisotropic spectrum. The average momentum
  scattering time in this scheme is also comparable to that  
  from the real band structure.

\subsection{HSR in intrinsic GaAs}
In this part, we investigate the HSR in intrinsic
case. Our discussion is based on two physical
quantities, i.e., the quasi-spin  polarization and the spin
polarization. The former describes the population difference between
the two HH and LH quasi-spin bands, defined as 
\begin{equation}
P^p=\sum_{\eta\bf k}(\rho^h_{\eta\bf k}(1,1)-\rho^h_{\eta\bf
  k}(2,2))/n_h,
\label{eq13}
\end{equation}
with $n_h$ being the hole density. The latter is calculated as the spin
polarization along the [001]-direction
\begin{equation}
P^s=\sum_{\eta\bf
k}{\rm Tr}(\rho_{\eta\bf k}^cJ_z)/n_h,
\label{eq14}
\end{equation}
which reflects
the optical orientation signal in experiment.
\begin{figure}[bth]
\includegraphics[width=6.5cm]{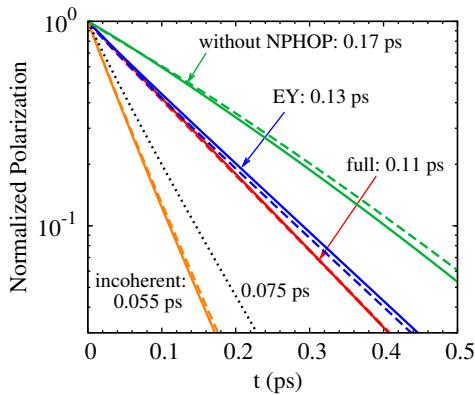}
\caption{(Color online) Temporal evolution of the normalized quasi-spin
  polarization (dashed curves) and spin polarization (solid ones) at room
  temperature with the hole density $n_h=10^{17}$~cm$^{-3}$ in
  intrinsic GaAs. The curves labelled 
as ``full'' (``incoherent'') are obtained from the full
  calculation with (without) the intraband coherence, while the ones
 with ``EY'' show the results without the intraband
  precession term, corresponding to the contribution from the EY
  mechanism. The results from the calculation including the intraband
  coherence but without the nonpolar
  hole-optical-phonon scattering are plotted as the curves denoted as
  ``without NPHOP''. The dotted curve represents
  the spin polarization from the calculation scheme used in
  Ref.\,\onlinecite{krauss}, i.e., without the nonpolar
  hole-optical-phonon scattering and intraband coherence. The
  relaxation times are given correspondingly.
} 
\label{fig2}
\end{figure}

In our computation, we take the initial state from
the optical orientation due to the pump pulse with the small polarization
$P^{\rm opt}=(I^+-I^-)/(I^++I^-)=2$\,\%,\cite{arg_p} where $I^+$ and $I^-$ are the intensities of
the $\sigma^+$- and $\sigma^-$-polarized light. In that case,
one sets the electron density matrices
as the Fermi distribution with the spin polarization 1\,\% at the lattice
temperature and keeps it unchanged, by taking into account of the fact
that the electron spin (momentum) relaxation 
time is much longer (shorter) than the time scale of the HSR.
The initial hole density matrices are also set to obey the Fermi
distribution in the collinear spin space to
describe the optically excited condition. By solving the
KSBEs, one obtains the temporal evolution of the
spin polarization $P^s$ (quasi-spin one $P^p$), from which the
spin (quasi-spin) relaxation time is extracted.

We first take the hole density
$n_h=10^{17}$~cm$^{-3}$. Figure\,\ref{fig2} 
illustrates the temporal evolution 
of the spin polarization (normalized by the
value at $t=0$) from the full calculation at room temperature as
the red solid curve (labelled as ``full''). One can see that the spin polarization
decays exponentially with the HSR time $\tau_{\rm tot}\approx 0.11$~ps,
which agrees perfectly well with the experimental value of the 
HSR time, $0.11\pm$10~\%~ps.\cite{hilton} By removing the 
coherent term from the KSBEs, one switches off the DP mechanism
  and obtains the results solely
due to the EY mechanism as the blue solid curve. The
HSR time $\tau_{\rm EY}$ in this case is about 0.13~ps. Since the
EY mechanism is irrelevant to  the spin precession, from
$\tau_{\rm tot}^{-1}=\tau_{\rm EY}^{-1}+\tau_{\rm DP}^{-1}$, one extracts
the spin lifetime due to the DP mechanism $\tau_{\rm
  DP}\approx0.72$~ps, which is much longer than that of the EY
mechanism. Therefore, the EY mechanism is dominant in this
case. Moreover, we find that the relaxation time of the
quasi-spin polarization (red dashed curve labelled with ``full'') 
is very close to that of the spin polarization.

\begin{figure}[bth]
\includegraphics[width=6.5cm]{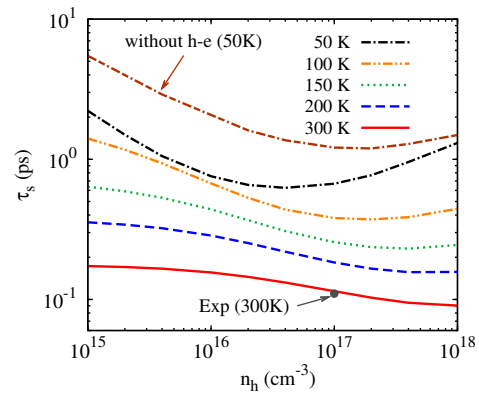}
\caption{(Color online) HSR time from the full calculation as
  function of the hole density $n_h$ in intrinsic GaAs. 
 The result from the calculation without the hole-electron
    scattering at 50~K is labeled as ``without h-e''.
  The dot represents the
  experiment data at room temperature with the photo-excitation density
  $n_h=10^{17}$~cm$^{-3}$ (Ref.\,\onlinecite{hilton}). In the
  calculation, the electron density $n_e=n_h$.}
\label{fig3}
\end{figure}

In the figure, we also plot the results from the calculation without
the intraband coherence as
orange curves (labelled as ``incoherent''), 
where the non-diagonal elements $\rho_{\eta\bf
  k}(1,2)$ are artificially set to be zero during the calculation as done in
Ref.\,\onlinecite{krauss}. Compared with the results labelled with
  ``EY'', one immediately finds that this procedure
markedly exaggerates the contribution of the EY mechanism, (The
DP mechanism vanishes in this scheme obviously.) which
clearly shows the problem of the missing of the intraband coherence. 
The reason lies in the fact that the spin-conserving scattering 
becomes the spin-flip one once the non-diagonal elements of
the density matrices in helix representation are
neglected.\cite{wu4}
In Ref.\,\onlinecite{krauss}, the authors argued  the disregarding of the
intraband coherence by arguing
that the optically driven  coherence between single particle states 
would be very small and the
SOC could not drive any coherence further during the time evolution of
the density matrices in the helix
representation (i.e., under the basis called ``intelligent basis'' there).
However, this is actually incorrect. Since the SOC 
information is transferred into the wave functions of the eigenstates
in the helix representation, any spin-conserving scattering events
can induce the intraband and/or interband coherence.\cite{wu4}
Therefore, the same HSR time is obtained if one
chooses the helix initial
condition, corresponding to the initialization without any
non-diagonal elements of the density matrices at $t=0$. 
In the remaining part of this paper, all the results are obtained from the
calculation including the intraband coherence.

The green curves (labelled as ``without NPHOP'') 
in Fig.\,\ref{fig2} obtained from the calculation without the
nonpolar hole-optical-phonon scattering give the HSR time
around 0.17~ps, much longer than that from the full calculation. This
clearly shows the importance of the nonpolar hole-optical-phonon
scattering in the HSR. It is noted
that this scattering is also missed in Ref.\,\onlinecite{krauss}. For
comparison, we further plot the spin 
polarization from the calculation without the nonpolar
hole-optical-phonon scattering and the intraband coherence
as black dotted curve, which corresponds to the calculation scheme in
Ref.\,\onlinecite{krauss}. It is seen that the missed terms have marked 
contribution to the HSR.

Since the direct relation between the
EY mechanism and the scattering strength, the HSR time due to the EY
mechanism decreases (increases) when the scattering becomes
stronger (weaker). To show the influence of the scattering more clearly,
we vary the density and temperature.
The density dependence of the
HSR time is plotted in
Fig.\,\ref{fig3}. One can see that the HSR time at room temperature
monotonically decreases with 
increasing photo-excitation density. Since the electrons
and holes are both in the nondegenerate regime, the Coulomb
scattering rate becomes larger as the density increases, according to
the estimation of the average Coulomb scattering rate
$(\tau_p^{C})^{-1}\propto n_h/T^{3/2}$.\cite{giulianni} As a result, the
HSR time decreases. One can also see that the
HSR time markedly increases with decreasing temperature,
due to the suppression of the hole-phonon scattering.

More interestingly, we predict a valley in the density dependence
of the HSR time at low temperature. The minimum occurs
at $n_h=n_e=4\times 10^{16}$~cm$^{-3}$ 
(corresponding to the Fermi
temperatures $T_F^e=74$~K for electrons and $T_F^h=9$~K for holes)
at $T=50$~K and $n_h=n_e=2\times 10^{17}$~cm$^{-3}$ ($T_F^e=216$~K and
$T_F^h=27$~K) at $T=100$~K. The agreement between $T_F^e$ at the
valley and the lattice temperature $T$ reveals that 
the valley results from the different density dependence of the
hole-electron scattering in the degenerate and nondegenerate
limits. As mentioned above, the HSR time decreases with
increasing density in the nondegenerate limit. However, for high
densities, electrons first enter into the degenerate
regime. Therefore the hole-electron scattering strength 
is significantly decreased due to the Pauli-blocking of electrons.
As a result, the HSR time due to the EY
 mechanism increases in this regime, and finally results in the
  valley at the crossover between the degenerate and nondegenerate
  regimes of electrons. In the figure, we also plot the result from the
calculation without the hole-electron scattering at 50~K. It is clear
to see that the valley at $n_h=n_e=4\times 10^{16}$~cm$^{-3}$ 
disappears, as expected. Instead, one finds 
that another valley shows up at
    $n_h=n_e=2\times 10^{17}$~cm$^{-3}$, where $T_F^h$ (27~K) is
  comparable to the lattice temperature $T$ (50~K). This means that
 holes also enter the degenerate regime at such
a high density and the hole-hole scattering rate is then given by
$(\tau_p^{C})^{-1}\propto T^2/n_h^{2/3}$.\cite{giulianni}
  Therefore, the HSR time increases with increasing the
density for $n_h=n_e>2\times 10^{17}$~cm$^{-3}$.
One notices that the Coulomb scattering can also induce a
peak (instead of valley) in the density dependence of the electron spin relaxation time
limited by the DP mechanism, first
predicted by Jiang and Wu\cite{wu6} and realized experimentally by Krau\ss\
{\em et al.}.\cite{krauss2,shen}

\subsection {HSR in $p$-type GaAs}
In this part, we study the HSR in $p$-type GaAs. We
still apply the collinear initial spin polarization
$P^s=2$~\%. As the photo-excited
carrier density is much smaller compared to the doping density, we take
$n_h=n_i$ and neglect the hole-electron scattering in the
computation.

\begin{figure}[bth]
\includegraphics[width=6.5cm]{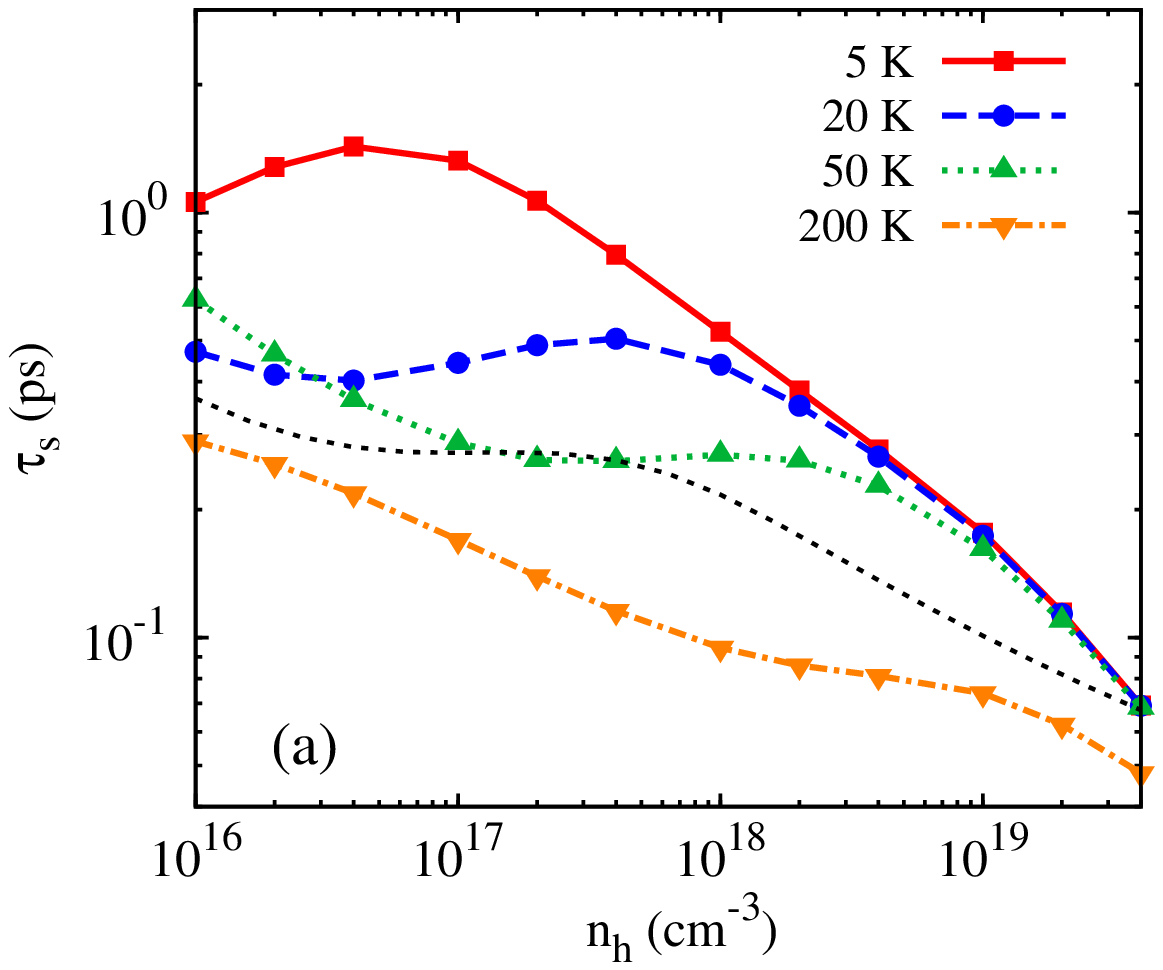}
\includegraphics[width=6.5cm]{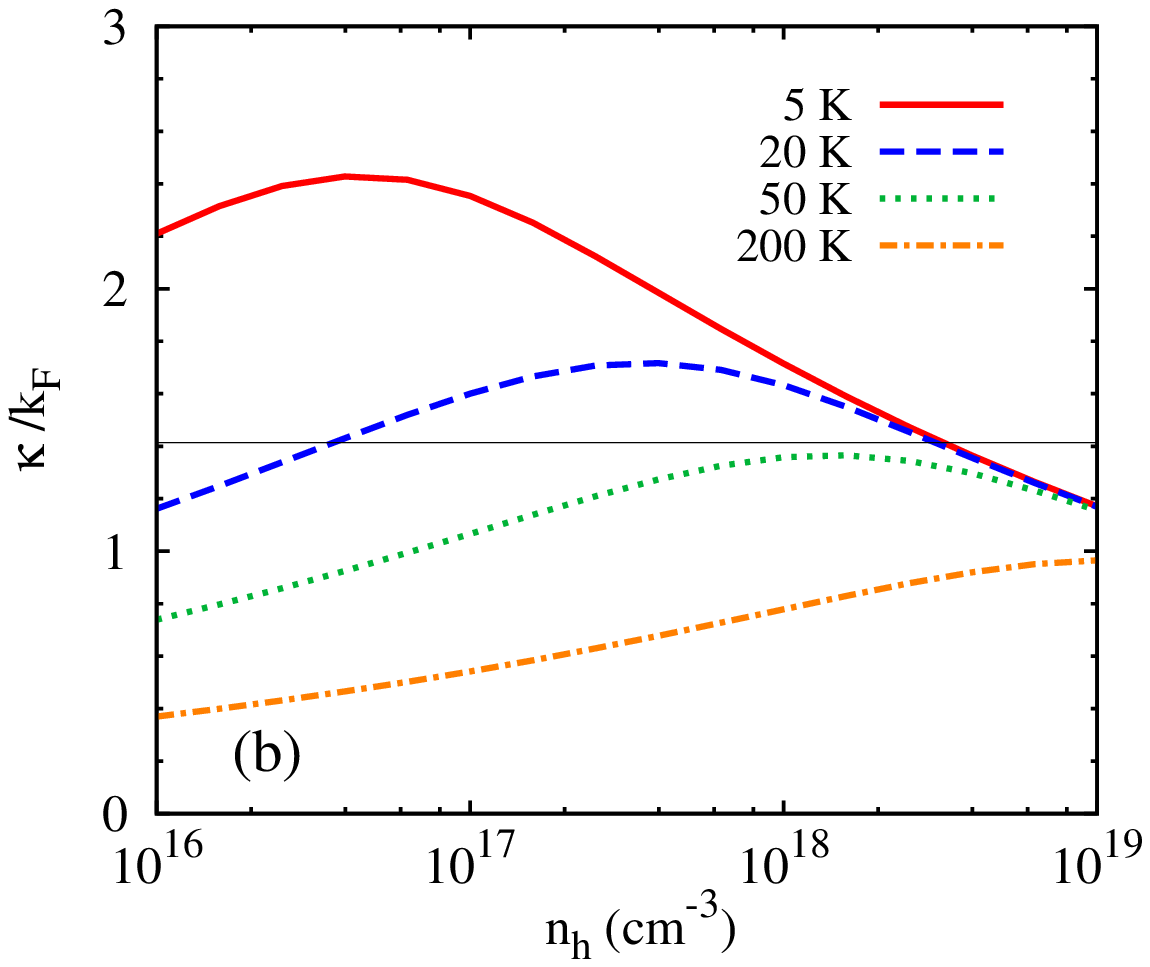}
\caption{(Color online) (a) HSR time from the full calculation as
  function of the hole density $n_h$ in $p$-type GaAs. We take the
  impurity density $n_i=n_h$. The hole-impurity scattering time
    $\tau_p^{hi}$ at 20~K is plotted as the black dotted curve without symbol.
  (b) The ratio of
  the screening constant to the 
  Fermi wave vector, $\kappa/k_F$, as function of the hole
  density. The solid line illustrates the level
  $\kappa=\sqrt{2}k_F$. } 
\label{fig4}
\end{figure}

\subsubsection{Density dependence}
We start from the density dependence of the HSR time.
The results from the full calculation 
are plotted in Fig.\,\ref{fig4}(a), which shows rich and intriguing
nonmonotonic features. At low temperature, e.g., at 5~K,
the HSR time first increases and then decreases 
with increasing hole density. As a result, a peak appears at
  $n_h=4\times 10^{16}$~cm$^{-3}$, where the corresponding Fermi
  temperature $T_F^h=9$~K is comparable to the lattice temperature. As the
temperature increases, the peak moves to the higher density
regime ($n_h=3\times 10^{17}$~cm$^{-3}$ with
$T_F^h=35$~K for 20~K and $n_h=10^{18}$~cm$^{-3}$ with
$T_F^h=79$~K for 50~K). Simultaneously, a valley
gradually appears in the low density regime. 
Although the peak also locates at the crossover between the degenerate
  and nondegenerate regime, the
underlying physics of the peak is again different from
the one in the electron spin relaxation time in intrinsic and
$n$-type materials.\cite{wu6} To explain the presence of the peak,
we turn to analyze the change of the scattering strength.
 Since the intraband splitting is too
small to affect the HSR for the hole states near the
zone-center, the HSR process is 
mainly determined by the EY mechanism instead of the DP
mechanism for low density case, especially at low temperature.

As a qualitative analysis for the case $n_i=n_h$, 
one can focus on the major scattering
mechanism, i.e., the hole-impurity scattering with the strength
estimated by 
$(\tau_p^{hi})^{-1}\sim m_{\rm eff}n_h\langle kV_q^2\rangle/(8\pi^2\hbar^3)\propto n_h\langle k
(q^2+\kappa^2)^{-2}\rangle$,
where $k$ comes from the density of states and $\langle ...\rangle$
stands for the average over the
distribution. $m_{\rm eff}$ represents the effective mass and $q$ is
the momentum exchange. In the degenerate 
regime, only holes on the Fermi surface contribute to the
HSR, therefore, one can estimate $k \approx k_F$ and $
q^2\approx 2 k_F^2$ with $k_F$ representing
the Fermi wave vector. Thus, one has
\begin{equation}
\tau_p^{hi}\propto [2+(\kappa/k_F)^2]^2.
\label{eq15}
\end{equation}
Figure\,\ref{fig4}(b) shows the ratio of the RPA screening constant
$\kappa$ to $k_F$. One finds that $\kappa/k_F$ decreases with
increasing density in the high density regime and $\tau_s\sim\tau_p^{hi}$ decreases also.
The density dependence of $\kappa/k_F$ can be easily understood once the
Thomas-Fermi screening ($\kappa\propto n_h^{1/6}$) is applied in the degenerate limit,\cite{haug}
which leads to $\kappa/k_F\propto n_h^{-1/6}$.

However, the situation is quite different in the nondegenerate regime.
In this limit, one has $\kappa\propto n_h^{1/2}T^{-1/2}$ according to the Debye-H\"uckel
screening,\cite{haug} and $q\sim k \sim
T^{1/2}$. For $\kappa\gg q$, one neglects the $q^2$ term in
  $\tau_p^{hi}$ and obtains
\begin{equation}
\tau_p^{hi}\propto (n_h\langle k \kappa^{-4}\rangle)^{-1}\propto
n_h T^{-5/2},
\label{eq16}
\end{equation}
which indicates that the HSR time increases with increasing
  density in this case. For $\kappa\ll q$, the screening constant is neglected, and
  $\tau_p^{hi}$ can be written as
\begin{equation}
\tau_p^{hi}\propto (n_h\langle k q^{-4}\rangle)^{-1}\propto
n_h^{-1} T^{3/2},
\label{eq17}
\end{equation}
which decreases with increasing density. 
In Fig.\,\ref{fig4}(a), we also plot the density dependence of 
  $\tau_p^{hi}$ at 20~K (the dotted curve without symbol), which
  agrees well with our discussion. It is clear to see that
  $\tau_p^{hi}$ is in the same order of magnitude as the HSR time
  $\tau_s$ as expected.

Now, the peak and valley in the HSR time in
Fig.\,\ref{fig4}(a) can be well understood. For example at 20~K, holes
lie in the nondegenerate regime and the screening constant is small in the
low density regime. As the density increases, the HSR time decreases
according to Eq.\,(\ref{eq17}).
Nevertheless, with further increase of the density, $\kappa$ can be larger
than $q$. Then the HSR time increases with density according to
  Eq.\,(\ref{eq16}). By further increasing density, the system 
  enters into the degenerate regime, and the HSR 
  time decreases again following Eq.\,(\ref{eq15}).
By comparing Fig.\,\ref{fig4} (a) and (b), we find the crossover
 between Eqs.\,(\ref{eq16}) and (\ref{eq17}) can still be qualitatively
estimated by taking $q^2\sim 2k_F^2$. At low temperature (5~K), the screening
  constant is always large in the nondegenerate regime of our
  investigation [see
  Fig.\,\ref{fig4}(b)], so the valley is invisible in Fig.\,\ref{fig4}(a). However,
  at high temperature (200~K), the screening is weak in both the
  degenerate and nondegenerate regimes and the HSR time
  monotonically decreases with increasing density.

\begin{figure}[bth]
\includegraphics[width=6.5cm]{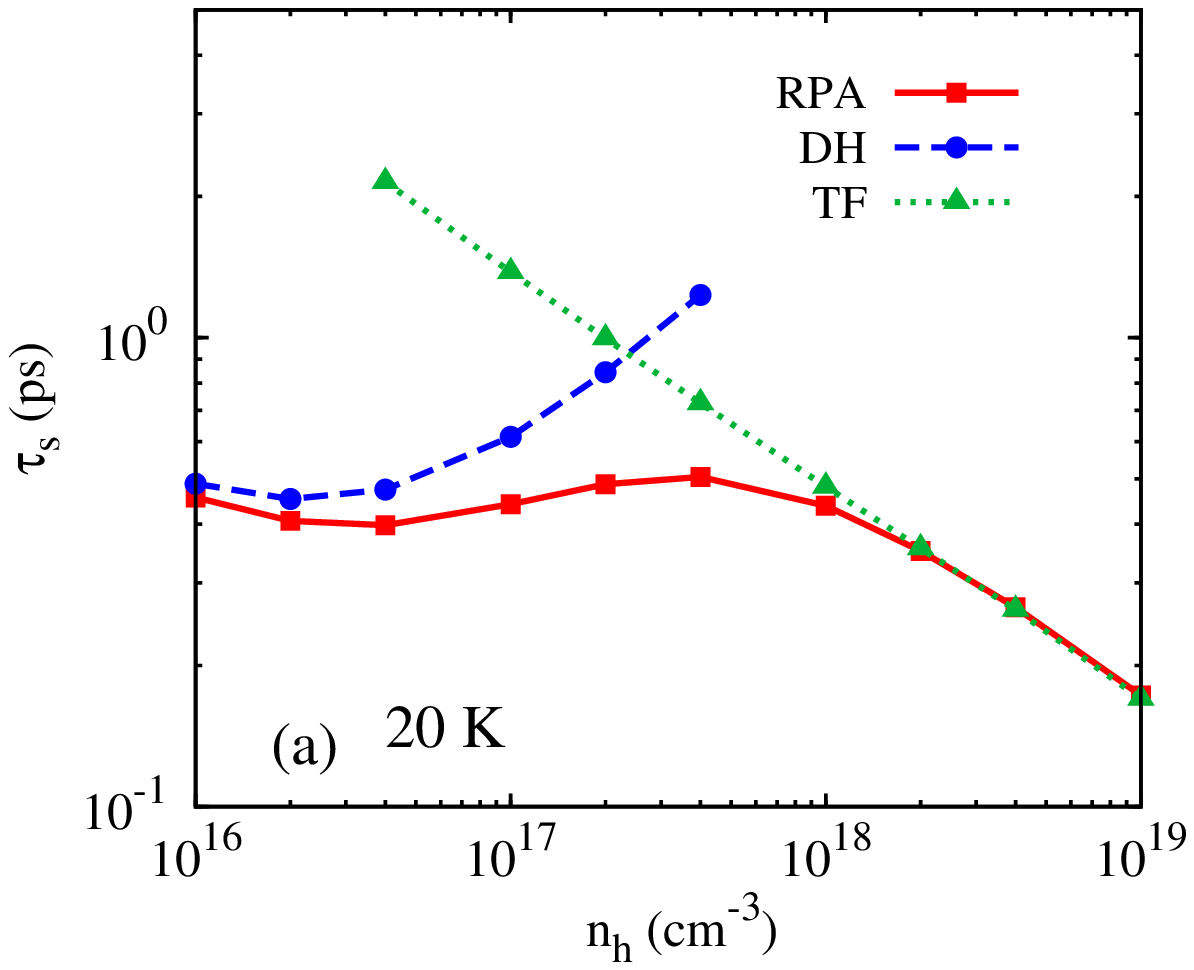}
\includegraphics[width=6.5cm]{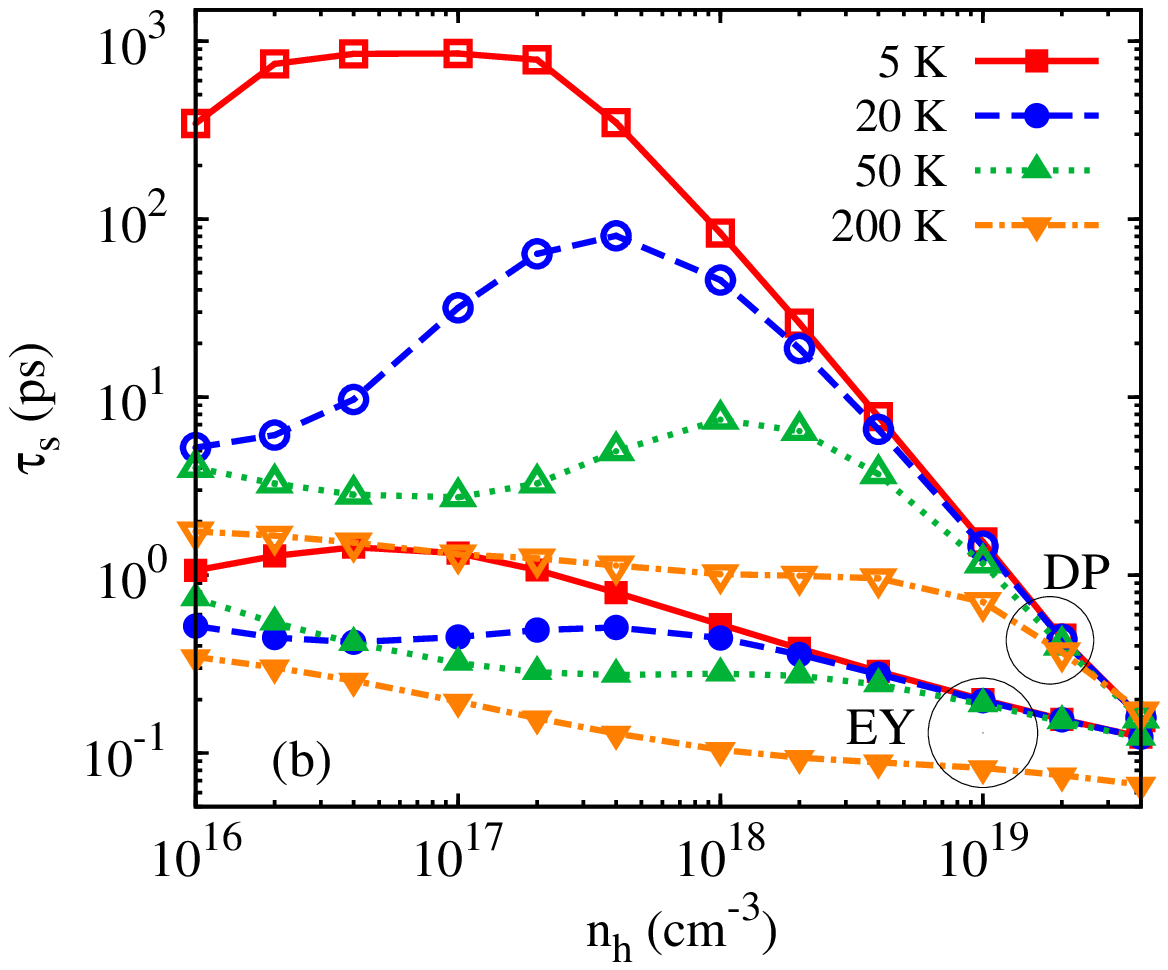}
\caption{(Color online) (a) Density dependence of the total HSR
  time in $p$-type GaAs at 20~K. In the calculation, the RPA (red solid
  curve), Thomas-Fermi (green dotted curve)
  and Debye-H\"uckel (blue dashed curve) screenings 
  are applied. (b) HSR
 times due to the EY mechanism (curves with
  solid symbols) and the DP
  mechanism (curves with open symbols) as function of the
  hole density. 
} 
\label{fig5}
\end{figure}

To elucidate the role of the screening more clearly, we plot the
results from different screenings at 20~K in Fig.\,\ref{fig5}(a). The
green dotted and blue dashed curves correspond to the results with the
Thomas-Fermi and Debye-H\"uckel screenings,
respectively. The total HSR time from the RPA 
screening is also plotted as the red solid curve. One can see
that the result from the RPA screening is consistent with that from
the Thomas-Fermi (Debye-H\"uckel) screening
in the high (low) density regime and the peak in the density 
dependence exactly occurs at the crossover between the degenerate and nondegenerate regimes.

In contrast, electrons in most $n$-type zinc-blende materials
are easier to enter into the degenerate regime and the screening constant
is usually small ($\kappa/k_F<\sqrt{2}$) thanks to the small
effective mass of the conduction band. Therefore, it is difficult to observe
the nonmonotonic effect of the screening during the electron spin relaxation.
However, Jiang and Wu suggested that the screening from the holes in
$p$-type materials can also give rise to observable effects on the electron
spin relaxation.\cite{wu6} In that case, the EY mechanism is always irrelevant and the
electron spin relaxation is dominated by the DP
mechanism in the low hole density regime. Since the spin relaxation
time due to the DP mechanism is inversely proportional to the
momentum scattering time in the strong scattering limit, the hole
density dependence of the electron spin relaxation time there is
opposite to that of the HSR time we predict here. Specifically, they 
found that the electron spin relaxation time due to the DP mechanism
first increases and then decreases as the hole 
density increases in the nondegenerate regime. After the holes enter
into the degenerate regime, the electron spin relaxation time
increases again.

From previous discussion, one notices that the HSR properties can be
well interpreted by the EY mechanism which suggests that the EY
mechanism is generally more important than the DP one. 
This can be clearly seen from Fig.\,\ref{fig5}(b), where 
the HSR times due to the EY and DP mechanisms are
separately plotted. Interestingly, we
find that the contribution of the DP mechanism can be comparable to that
of the EY mechanism in high doping regime.
To explain this behavior, we employ the relation
in the strong scattering limit, $\tau_{\rm DP}\sim 1/[\langle |{\bgreek
  \Omega}|^2-\Omega_z^2\rangle \tau_p^{hi}]$ with $\langle |{\bgreek 
  \Omega}|^2-\Omega_z^2\rangle$ representing the ensemble average of the effective
magnetic field (inhomogeneous broadening\cite{wu1}).
In the low density regime, the hole gas is
in the nondegenerate limit and the inhomogeneous broadening
is small and insensitive to the density. Hence the contribution of the
DP mechanism to the HSR
is negligible. However, holes are in the degenerate limit for high
density case and the inhomogeneous broadening increases rapidly
[$\propto k_F^6$, see Eq.\,(\ref{eq6}) also]
with increasing hole density, which makes
the DP mechanism markedly contribute to the HSR. 
By using Eq.\,(\ref{eq15}), one can easily obtain $\tau_{\rm DP}/\tau_{\rm EY}\propto n_h^{-2/3}$ for
$\kappa/k_F\gg \sqrt{2}$ and $\tau_{\rm DP}/\tau_{\rm EY}\propto n_h^{-2}$ for
$\kappa/k_F\ll\sqrt{2}$. It is obvious that $\tau_{\rm
  DP}/\tau_{\rm EY}$ decreases with increasing density in both cases.

\begin{figure}[bth]
\includegraphics[width=6.5cm]{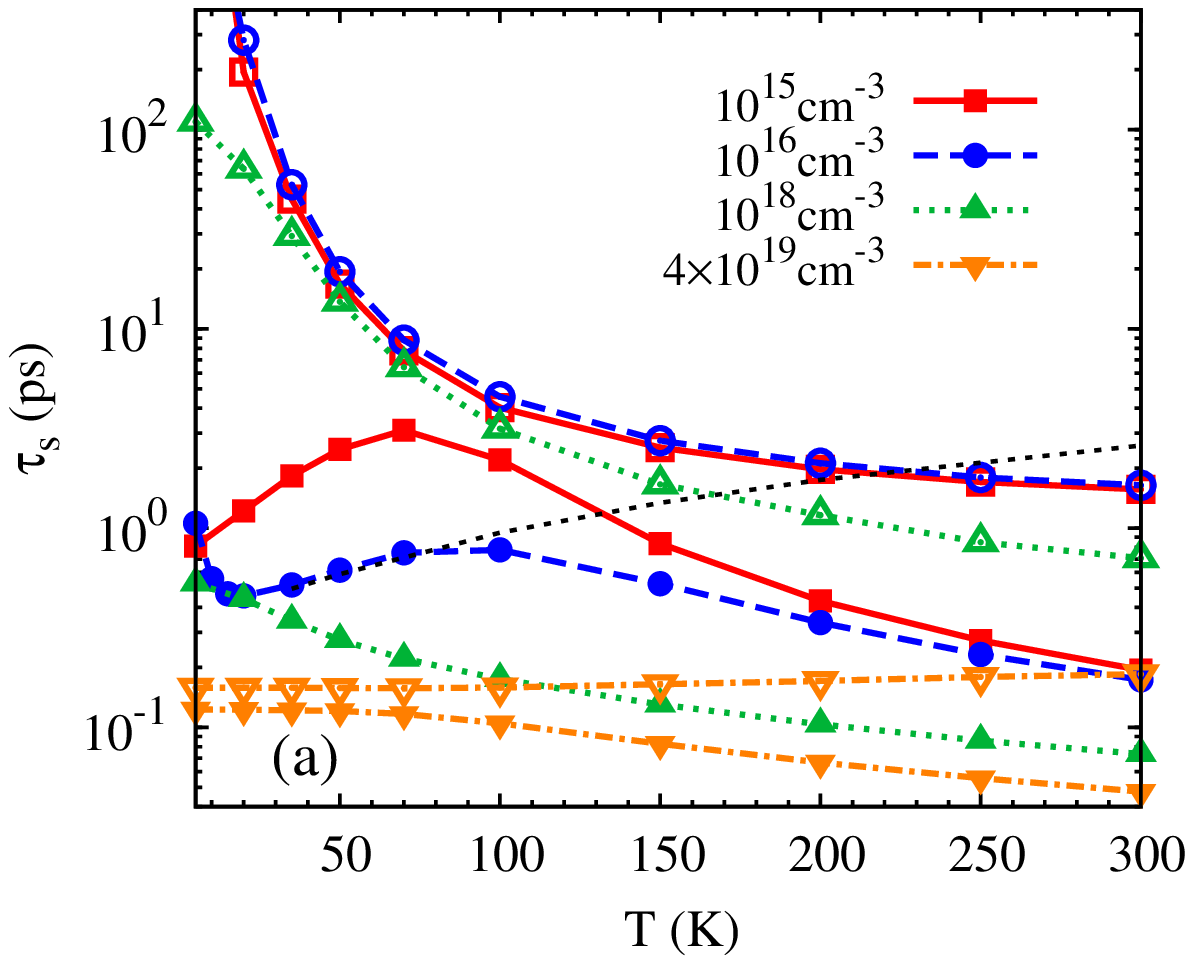}
\includegraphics[width=6.5cm]{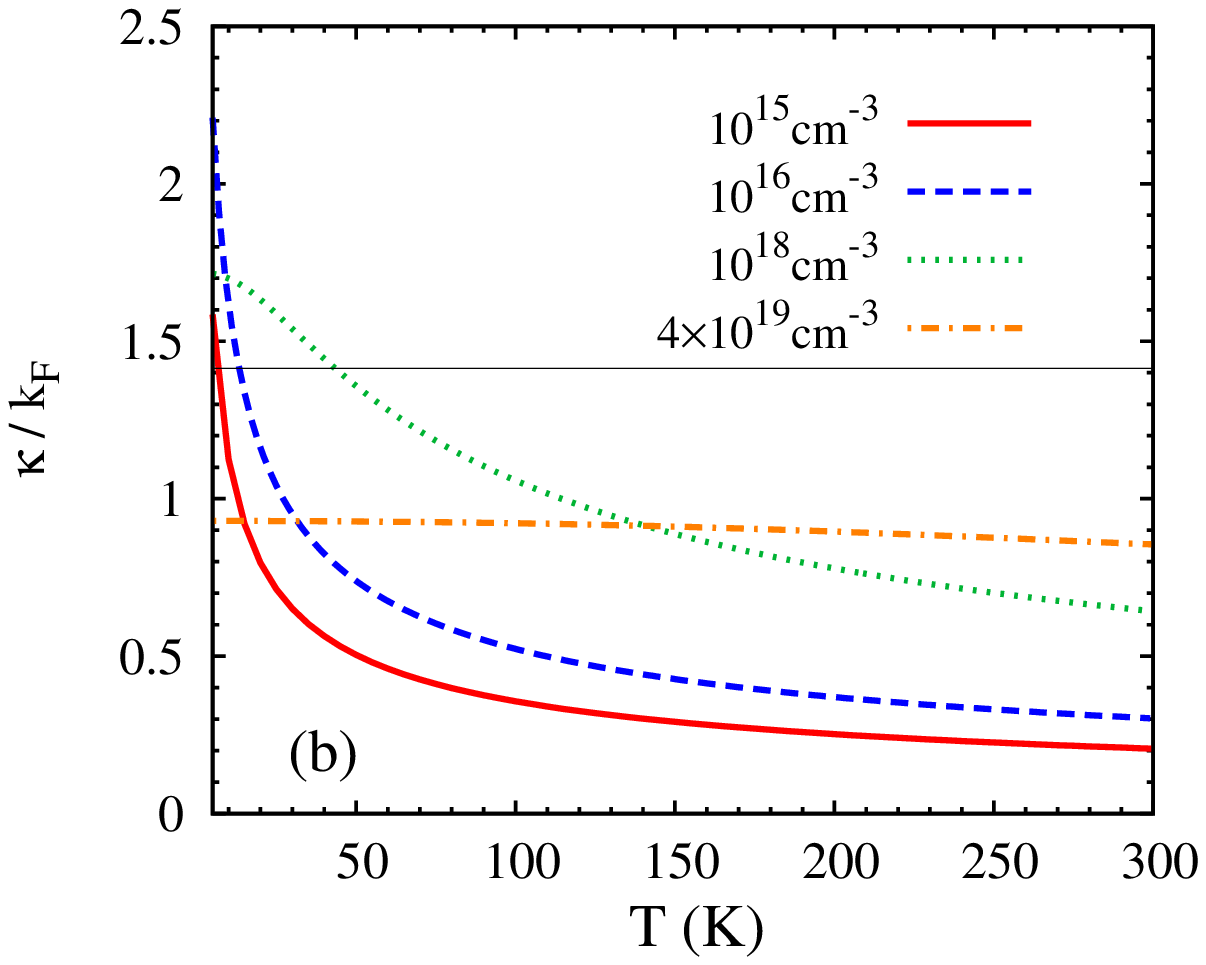}
\caption{(Color online) (a) HSR times due to the EY
  (curves with solid symbols)
  and DP (curves with open symbols) mechanisms as function of the
  temperature in $p$-type GaAs with $n_i=n_h$. The result
  without the hole-phonon scattering at $n_h=10^{16}$~cm$^{-3}$ is
  plotted as dotted curve without symbol. (b) The ratio of the
  screening constant to the Fermi wave vector as function of the
  temperature. The level $\kappa=\sqrt{2}k_F$ is illustrated by a
  solid line.
} 
\label{fig6}
\end{figure}

\subsubsection {Temperature dependence}
Similarly, one also finds the nonmonotonic temperature dependence of the
HSR time.
In Fig.\,\ref{fig6}(a), we plot the results of four typical densities
$n_h=10^{15}$, $10^{16}$, $10^{18}$ and $4\times
10^{19}$~cm$^{-3}$. One notices that the temperature dependence of the
HSR time is also qualitatively determined
by the EY mechanism (curves with solid 
symbols). The HSR time from this mechanism
first decreases and then 
increases with increasing temperature for $n_h=10^{16}$~cm$^{-3}$, and the
minimum reaches around 15~K. Since the hole-phonon scattering is
rather weak in this regime, this feature just reflects the important
role of the screening in the hole-impurity
scattering. Figure\,\ref{fig6}(b) illustrates the ratio of the
screening constant to the Fermi wave vector, where
the crossover ($\kappa\approx \sqrt{2}k_F$) occurs just around
15~K. Since holes are always nondegenerate in this case
(even at 5~K, see Fig.\,\ref{fig4}), the HSR time decreases (increases)
with increasing temperature below (above) 15~K, according to Eq.\,(\ref{eq16})
[Eq.\,(\ref{eq17})]. Moreover, we find that the HSR time decreases
again when the temperature further increases, because of the
enhancement of the hole-phonon scattering. To
demonstrate this picture, we remove the hole-phonon scattering
from the KSBEs and find that the HSR time due to the EY mechanism
monotonically increases above 15~K, as expected
(shown as dotted curve without symbol). For $n_h=10^{15}$~cm$^{-3}$, only the peak is visible
while the valley is absent as the curve with solid squares shown.
The reason is that the screening is weak for this density [see
Fig.\,\ref{fig6}(b)]. However, for $n_h=10^{18}$ and $4\times
10^{19}$~cm$^{-3}$, holes are 
in the degenerate regime at low temperature, hence the temperature
dependence of the HSR time is determined by $\kappa/k_F$ from
Eq.\,(\ref{eq15}). At high temperature regime, the HSR time is also limited
by the hole-phonon scattering for these densities. Another important information
one can get from Fig.\,\ref{fig6}(a) is
that the HSR time due to the DP mechanism is comparable to the
one due to the EY mechanism around 80~K for $n_h=10^{15}$~cm$^{-3}$. For the high doping case with
$n_h=4\times 10^{19}$~cm$^{-3}$, the important role of the DP
mechanism at low temperature can also be clearly seen [see also
  Fig.\,\ref{fig5}(b)].

\begin{figure}[bth]
\includegraphics[width=8cm]{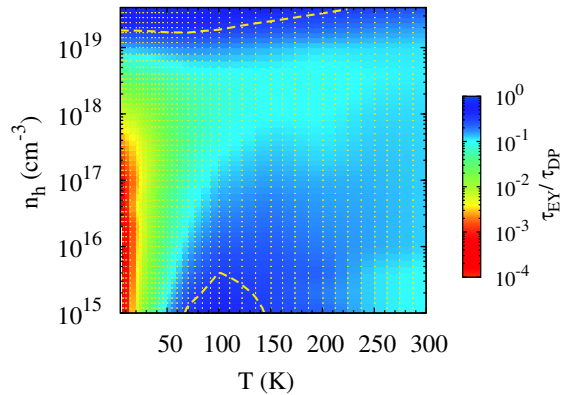}
\caption{(Color online) Ratio of the HSR time due to the EY mechanism to
  that due to the DP mechanism, $\tau_{\rm EY}/\tau_{\rm DP}$, as
  function of the doping density and temperature in
  $p$-type GaAs with $n_i=n_h$. The yellow dashed curves represent the
  borders with $\tau_{\rm EY}/\tau_{\rm DP}=1/3$ 
} 
\label{fig7}
\end{figure}

\subsubsection {Relative contribution of the DP and EY mechanisms}
For thorough understanding of the relative contribution of the DP and EY mechanisms,
we plot the ratio of the HSR times due to the EY and DP mechanisms
$\tau_{\rm EY}/\tau_{\rm DP}$ as function of the density and
temperature in Fig.\,\ref{fig7}. It is clear to see that the HSR time
is dominated by the EY mechanism at low temperature (below 40~K) upto
$10^{19}$~cm$^{-3}$. This is because that holes mainly occupy the
states with small wave vectors and experience a small effective Dresselhaus
field. In the high density regime, the increase of the inhomogeneous
broadening makes the DP contribution significantly enhanced and
even comparable to the EY one as discussed above. Another interesting regime lies
in the low density regime at moderate temperature, where the
contribution of the DP mechanism can also be comparable to that of
the EY mechanism. This originates from the reduction of the EY contribution
due to the relatively weak hole-impurity scattering according to
Eq.\,(\ref{eq17}) [see Fig.\,\ref{fig6}(a) also]. However, this effect
is suppressed by the hole-phonon scattering at higher temperature, where the EY
mechanism again is more efficient than the DP one. In the
figure, we also plot the borders with $\tau_{\rm EY}/\tau_{\rm
  DP}=1/3$ as yellow dashed curves.

Finally, we should point out that the Bir-Aronov-Pikus mechanism\cite{BAP}
is neglected in our computation even in the intrinsic case. The reason lies
in the fact that the spin-flip process due to the electron-hole
exchange interaction in intrinsic GaAs occurs in the time scale of
nanosecond,\cite{wu6} therefore this mechanism is irrelevant to the ultrafast
spin relaxation of the hole system. 

\section{Conclusion}\label{conclusion}
In conclusion, we have investigated the HSR from the fully
microscopic KSBEs in intrinsic and $p$-type bulk GaAs. We analyze the 
valence-band structure by considering the anisotropic property and the
Dresselhaus spin-orbit coupling. We find that
the degeneracy of the HH band and that of the LH
band are both lifted by intraband splittings. The DP mechanism
associated with the intraband 
precessions and the EY mechanism due to the direct spin-flip scattering
are then explicitly studied, with the intraband coherence
included. Our result of 
intrinsic GaAs shows good agreement with the experiment data at
room temperature, where the EY mechanism is demonstrated to be the
dominant spin relaxation mechanism due to the strong scattering
process. We also show that the approach without the intraband
coherence used in the literature 
is inadequate in accounting for the HSR and the
nonpolar hole-optical-phonon, missed in
the previous theoretical work, is important to the HSR. The
temperature can markedly affect the
HSR time by changing the strength of the hole-phonon
scattering. At low temperature, 
we predict a valley due to the Coulomb scattering in the
HSR time at the crossover between the 
degenerate and the nondegenerate regimes of the electrons. 
In $p$-type GaAs, we find that the HSR time
depends on density and temperature nonmonotonically. For the density
dependence, the HSR time presents a peak at the crossover between
the degenerate and nondegenerate regimes of the holes,
resulting from the different features of the screening in the degenerate
and nondegenerate limits. In the nondegenerate regime, we
predict a valley, from the competition between the screening constant
and the momentum exchange, in the density dependence of the HSR
time. We also find that the HSR time monotonically decreases with
increasing density at high
temperature, thanks to the small screening constant. For the
temperature dependence, we predict a valley in the low temperature
regime, which also reflects the role of the screening. In the high
temperature regime, the hole-phonon scattering can markedly
contribute to the HSR and make the HSR time decrease with increasing
temperature. Moreover, we find that the contribution of the DP mechanism can be
comparable with that of the EY one in the high density regime at low
temperature and in the low density regime at moderate temperature, even
though the EY mechanism is usually the major mechanism in the HSR.

\begin{acknowledgments}
This work was supported by the Natural Science Foundation of China
under Grant No.~10725417, the National Basic Research Program of
China under Grant No.~2006CB922005 and the Knowledge Innovation
Project of Chinese Academy of Sciences. We would like to thank
  E. L. Ivchenko for helpful discussion.
One of the authors (K.S.) would also
like to thank J.H. Jiang for valuable discussions.
\end{acknowledgments}

\appendix
\section{Detail of intraband splitting}\label{appen}

The coefficients in Eq.\,(\ref{eq4}) can be expressed as
\begin{eqnarray}
a_1^{h/l}&=&(E^{h/l}-G)/[{(E^{h/l}-G)^2+|H|^2+|I|^2}]^{\frac{1}{2}},\\
b_1^{h/l}&=&H^\ast/[(E^{h/l}-G)^2+|H|^2+|I|^2]^{\frac{1}{2}},\\
c_1^{h/l}&=&I^\ast/[(E^{h/l}-G)^2+|H|^2+|I|^2]^{\frac{1}{2}},
\label{eqA1_3}
\end{eqnarray}
and those in Eq.\,(\ref{eq4-5}) are given by $a_2^{h/l}=a_1^{h/l}$,
$b_2^{h/l}=-(b_1^{h/l})^\ast$ and
$c_2^{h/l}=(c_1^{h/l})^\ast$.

The effective magnetic field in Eq.\,(\ref{eq6}) can be written as
\begin{eqnarray}
  \nonumber
  \Omega^e_x&=&h_x{\rm
    Re}(\sqrt{3}a_2c_2-b_2^2+c_2^2)-h_z{\rm Re}(b_2c_2)\\
  &&\mbox{}+h_y{\rm Im}(\sqrt{3}a_2c_2-b_2^2-c_2^2),\\
\nonumber
  \Omega^e_y&=&-h_x{\rm
    Im}(\sqrt{3}a_2c_2-b_2^2+c_2^2)+h_z{\rm Im}(b_2c_2)\\
  &&\mbox{}+h_y{\rm Re}(\sqrt{3}a_2c_2-b_2^2-c_2^2),\\
\nonumber
  \Omega^e_z&=&h_x{\rm Re}(\sqrt{3}a_2b_1+2b_1c_2)+h_y{\rm Im}(\sqrt{3}a_1b_1-2b_2c_1)\\
  &&\mbox{}+h_z(3a_1^2+|b_1|^2-|c_1|^2)/2,
\label{eqA4_5}
\end{eqnarray}
where the labels ``$h/l$'' are neglected for short.

\section{Deformation potential matrix of optical phonons}\label{nonpolar}
The $6\times 6$ deformation potential matrix of optical phonons
for valence bands reads\cite{scholz} 
\begin{equation}
{\bar D}_{\lambda{\bf q}}=\frac{d_0}{a}\left(
\begin{array}{cc}
T_{\lambda\bf q}& P^\dag_{\lambda\bf q}\\
P_{\lambda\bf q}& 0_{2\times 2}
\end{array}
\right),
\label{eqB1}
\end{equation}
with
\begin{eqnarray}
\hspace{-1cm}T_{\lambda\bf q}&=&\left(
\begin{array}{cccc}
0& \delta_\lambda^+ & i\delta_\lambda^z &0 \\
\delta_\lambda^- &0 &0 &i\delta_\lambda^z  \\
-i\delta_\lambda^z &0 &0 &-\delta_\lambda^+ \\
0 &-i\delta_\lambda^z &-\delta_\lambda^- &0 
\end{array}
\right),\\
P_{\lambda\bf q}&=&\left(
\begin{array}{cccc}
-i\sqrt{\tfrac{1}{2}}\delta_\lambda^-& 0
&i\sqrt{\tfrac{3}{2}}\delta_\lambda^+ &-\sqrt{2}\delta_\lambda^z\\ 
\sqrt{2}\delta_\lambda^z&-i\sqrt{\tfrac{3}{2}}\delta_\lambda^- &0
&i\sqrt{\tfrac{1}{2}}\delta_\lambda^+
\end{array}
\right),
\label{eqB2_3}
\end{eqnarray}
where $\delta_\lambda^\pm=\delta_{\lambda x}\pm i\delta_{\lambda y}$
and $\delta_\lambda^z=\delta_{\lambda z}$. $d_0$
and $a$ are optical deformation potential and lattice constant,
respectively.


\begin{thebibliography}{0}
\bibitem{arg_p} In the small polarization case, the HSR
  time is insensitive to the polarization according to our computation.\cite{wu}
\bibitem{com_spliting} Even in the lowest density and temperature
  limit in our computation, i.e., $n_h=10^{15}$~cm$^3$ and $T=5$~K,
  the spin relaxation are mainly determined by the heavy holes
  with energies around 1.5~meV. The interband splitting there is about 10~meV, 
  much larger than $k_BT\approx 0.5$~meV. Therefore, it is still
  reasonable to treat the LH and HH separately.
\end{thebibliography}
\end{document}